# Comparative Study of Planetary Atmospheric Uncertainties and Design Rules for Aerocapture Missions

Athul Pradeepkumar Girija [1***]

[1]*School of Aeronautics and Astronautics, Purdue University, West Lafayette, IN 47907, USA*

## ABSTRACT

Aerocapture uses atmospheric drag to decelerate spacecraft and achieve orbit insertion. One of the significant risks associated with aerocapture is the uncertainty in the atmospheric density, particularly for outer planets. The paper performs a comparative study of the atmospheric uncertainties and provides design rules for aerocapture missions. The atmospheres of Venus, Mars, and Titan are well-characterized for engineering purposes. At the altitude ranges relevant for aerocapture, the $3\sigma$ density variation is approximately ±30%, ±50%, ±30% for Venus, Mars, and Titan respectively. With no in-situ data, the atmospheres of Uranus and Neptune are not as well characterized as the other bodies. For both Uranus and Neptune, the GRAM suite provides a $3\sigma$ density variation of approximately ±30% for the relevant altitude ranges which is considered an optimistic estimate. Until in-situ data from an atmospheric probe becomes available, a more conservative global min-max estimate is recommended to accommodate the worst-case scenario. The study presents a graphical method for selection of the optimal entry flight path angle when considering the atmospheric uncertainties to ensure the on-board guidance is given the best possible initial state for targeting the desired exit state post aerocapture.

***Keywords:*** Planetary atmosphere, GRAM, Uncertainties, Aerocapture

****** To whom correspondence should be addressed, E-mail: athulpg007@gmail.com

# I. INTRODUCTION

With the exception of Mercury, all planetary destinations in the Solar System and Saturn's moon Titan possess significant atmospheres [1]. From the hot thick Venusian $CO_2$ atmosphere to the cold icy $H_2$-He atmospheres of Uranus and Neptune, there exists great diversity in the physical structure and chemical composition of these atmospheric layers [2]. Measurements such as the noble gas abundances and isotopic ratios in these atmospheres are critical to our understanding of the origin, formation, and evolution of the Solar System [3]. The presence of atmosphere makes these destinations fundamentally more interesting than airless bodies, due to their ability to maintain a climate, induce weather phenomena such as storms and rainfall, and erosive process which shape these bodies [4, 5]. In addition to their scientific importance, the atmospheres are also of significant engineering interest for planetary missions. All planetary entry missions to date have utilized the drag provided by the atmosphere to slow down probes and landers [6, 7]. While it has never been flown, the related technique of aerocapture shown in Fig. 1 which uses atmospheric drag to decelerate spacecraft and achieve orbit insertion is of great interest for future missions across the Solar System [8, 9]. One of the significant risks with aerocapture is the uncertainty in the atmospheric density, particularly for outer planets whose atmospheres have no in-situ measurements as ground truth for the engineering models [10]. The paper uses the NASA Global Reference Atmospheric Model (GRAM) to perform a comparative study of the atmospheric uncertainties and provides design rules for aerocapture missions.

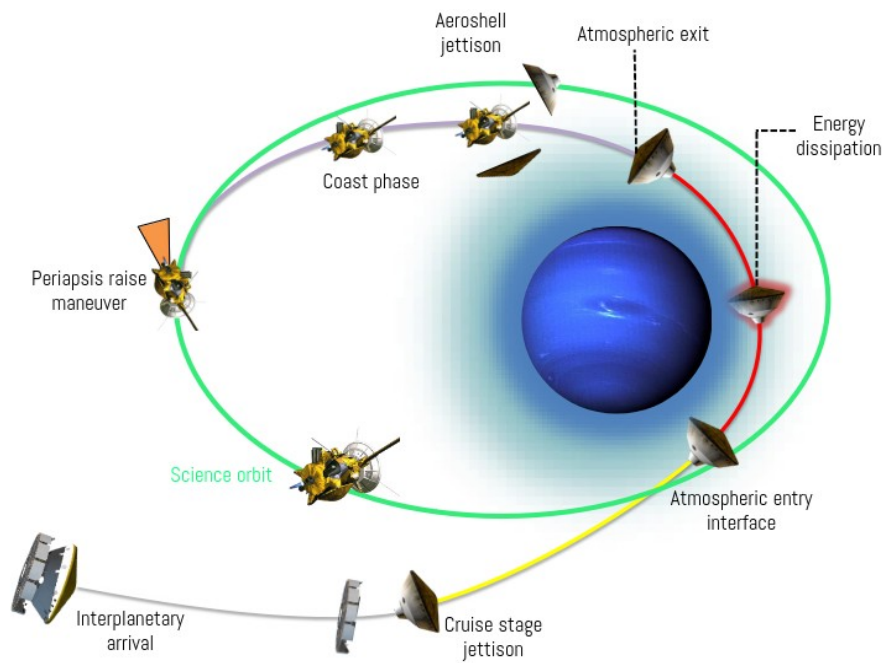

Figure 1.  Schematic illustration of the aerocapture maneuver.

## II. STRUCTURE AND CHEMICAL COMPOSITION

Figure 2 shows the extent and chemical composition of the atmospheres of which are of interest for aerocapture missions. The terrestrial planets Venus and Mars have well understood atmospheres that extend to about 120 km and is almost entirely composed of $CO_2$. Titan with its low gravity has a thick $N_2$ atmosphere that extends to nearly 1000 km above its surface. With in-situ measurements from from the Huygens Atmospheric Structure Instrument (HASI), Titan's atmospheric structure and composition ($CH_4$ mass fraction) is now well understood for engineering purposes [11]. Jupiter and Saturn are not considered in this study because their extreme entry speeds and harsh aero-thermal environments make aerocapture impractical [12]. The ice giant planets Uranus and Neptune have primarily $H_2$-He atmospheres with a small fraction of $CH_4$ which absorbs in the red, and gives them their distinctive blue-green color. The only measurements of the ice giant atmospheres are remote sensing observations from the Voyager flyby, and the radio science experiments using signals that passed through the atmosphere and then received on Earth. The lack of in-situ measurements make their atmospheres the least well constrained and has the largest uncertainties, posing a potential risk for aerocapture. The $CH_4$ mass fraction could also influence the aerothermodynamics and add uncertainty to the heating rates encountered. Until an atmospheric probe enters the Uranus atmosphere, potentially in the 2040s, our understanding of the ice giant atmospheres will remain relatively poor and with large uncertainties.

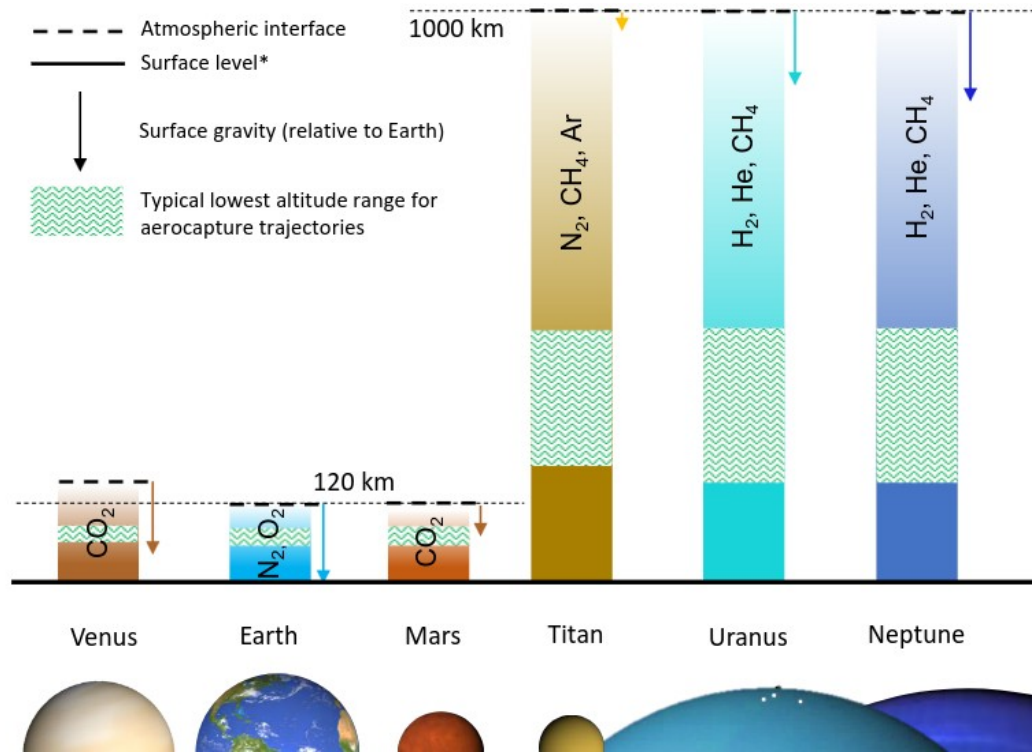

Figure 2. Extent and chemical composition of various planetary atmospheres.

## III. EFFECT OF ATMOSPHERIC UNCERTAINTIES

Aerocapture involves entry, atmospheric flight, and exit as shown in Figure 1. The density profile encountered during the atmospheric flight greatly affects the trajectory, and hence an understanding of the expected uncertainties in the density profile is of great importance to assess the risk it poses to a future mission. If the vehicle enters too shallow or encounters an atmosphere which is less dense than the expected minimum, spacecraft may exit the atmosphere without getting captured as shown in Figure 3. If the vehicle enters too steep, or the density is much higher than expected, the vehicle may bleed too much speed and fail to exit the atmosphere. Both of the above are are undesirable scenarios which will lead to complete loss of mission and hence adequate margins must be provided for the guidance system against these atmospheric uncertainties, in addition to delivery error and aerodynamic uncertainties [13]. The Theoretical Corridor Width (TCW) is a useful concept which quantifies the width of the corridor, and must be large enough to accommodate the delivery and atmospheric uncertainties, and also provide sufficient safety margin for mission success even in limiting scenarios (such as combination of shallow entry and thin atmosphere). NASA has developed the GRAM Suite as a unified tool, to provide engineering models for all planetary atmospheres [14]. The atmospheres of Mars and Venus are well-constrained for all engineering purposes, and the atmospheric uncertainties are quite low, while the ice giant atmospheres are significantly less constrained. The remainder of the paper uses the GRAM models to perform a quantitative study of the atmospheric uncertainties at various destinations.

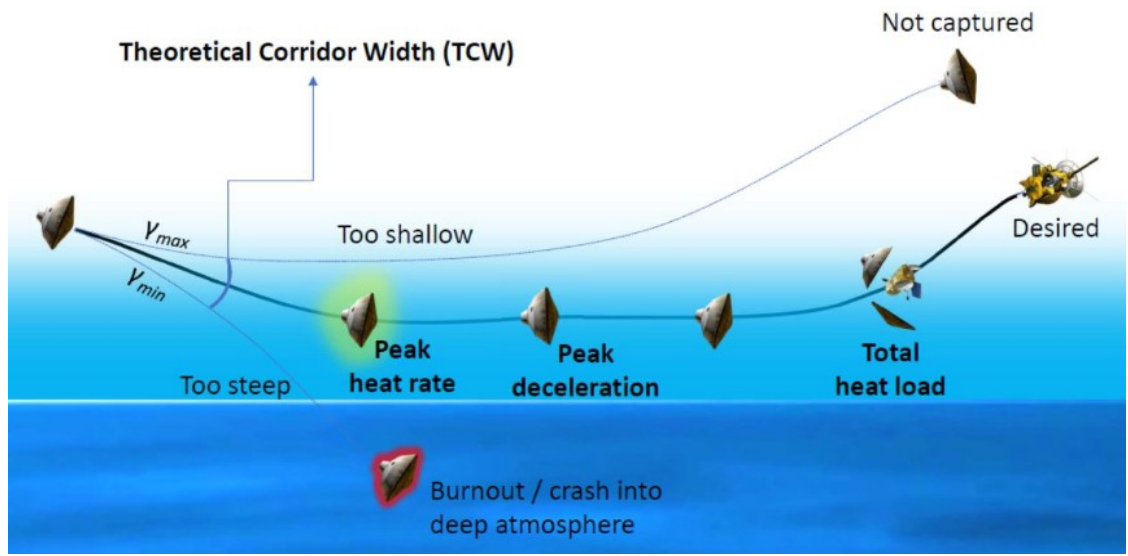

Figure 3. Illustration of the aerocapture theoretical corridor width.

## IV. VENUS

Venus is our closest planetary neighbor and aerocapture using its atmosphere has been shown to be feasible using both lift and drag modulation [15]. However, the large heating rates at Venus make lift modulation not attractive. Drag modulation with its lower heating rate particularly makes it attractive for small satellite orbit insertion, and has been extensively studied in the recent years in the context of low-cost missions [16, 17]. Drag modulation aerocapture at Venus has been proposed for inserting independent small satellites into low-circular orbits [18, 19], small satellites as part of New Frontiers or Flagship missions [20], and a future sample return mission from the Venusian cloud layers [21]. Due to the abundance of in-situ data from the Venera and Pioneer Venus entry probes, the Venusian atmospheric density from Venus-GRAM is practically well-characterized for all engineering purposes [22]. Venus-GRAM provides the average, low and high density ($1\sigma$) values. Figure 4 (left) shows the minimum, average, and maximum ($3\sigma$) density profiles as a function of altitude from Venus-GRAM. The shaded altitude band of 100–120 km is where most of the deceleration occurs for aerocapture at Venus, and hence the most relevant in terms of uncertainty quantification. Note that the results in Figure 4 are only for a particular location (latitude, longitude) and time (year, month, date). However, the results are expected to provide a reasonable estimate of the expected uncertainty for any location and time of year. Figure 4 (right) shows the percentage deviation ($3\sigma$) from the average as a function of altitude. Figure 4 shows that in the altitude range of 100–120 km relevant to aerocapture, the expected $3\sigma$ density variation is approximately ±30%. Venus aerocapture concepts should demonstrate success with these uncertainties accounted for.

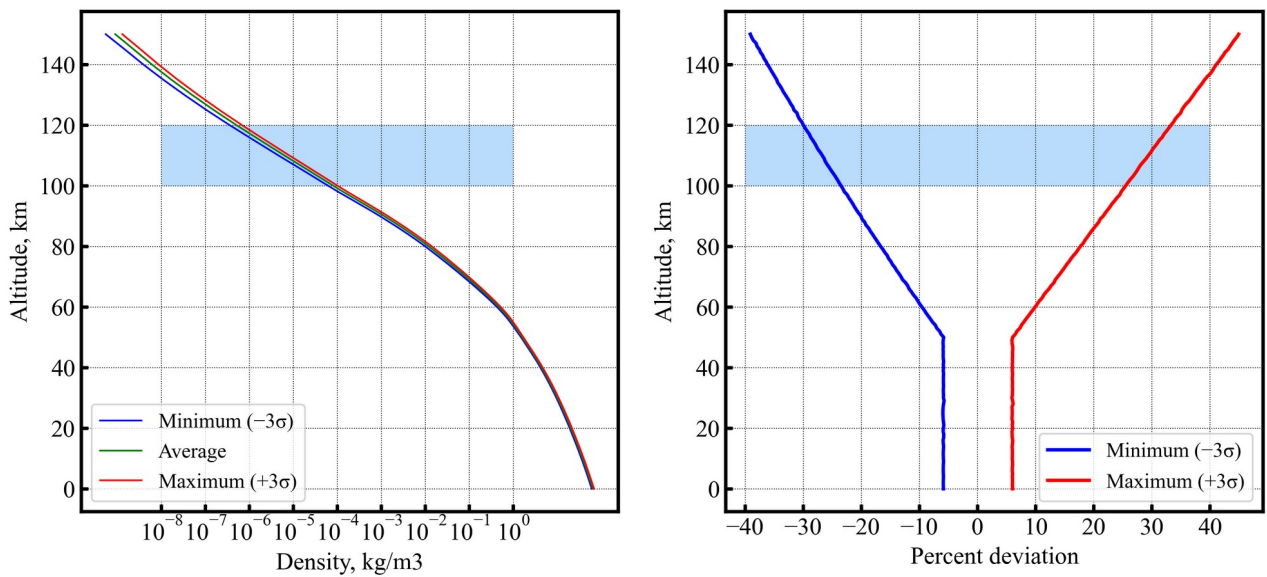

Figure 4. Density profiles from Venus-GRAM (left) and percent deviation from nominal (right).

## V. MARS

Mars has a relatively thin atmosphere compared to the Earth, but nevertheless is relevant for aerocapture and has been shown to provide small but still significant performance benefits [23]. The thinner atmosphere and the lower entry speed result in a relatively benign aero-thermal environment making it an attractive destination for a low-cost aerocapture technology demonstration [24, 25]. Drag modulation at Mars has been extensively studied in the recent years in the context of small low-cost planetary science missions [26]. Due to the plethora of lander and rover missions, the Martian atmosphere is also well understood, but also has relatively large seasonal variations compared to Venus and associated uncertainties, particularly in the thinner upper atmosphere. Figure 5 (left) shows the minimum, average, and maximum (3σ) density profiles as a function of altitude from a Mars-GRAM run. As with Venus, the results in Figure 5 are only for a particular location and time, but expected to be a representative estimate of the uncertainties. The shaded altitude band of 50–80 km is where most of the deceleration occurs for aerocapture at Mars. Figure 5 (right) shows the percentage deviation (3σ) from the average as a function of altitude. The uncertainty in the density profile increases with altitude. In the altitude range of 50–80 km relevant to aerocapture, the expected 3σ density variation is approximately ±50%. Compared to Venus, the low gravity and the extended atmosphere provide larger TCW at Mars (by a factor of 2), and hence larger atmospheric uncertainties can easily be accommodated. In particular, proposed concepts should be able to demonstrate mission success with adequate margin in two limiting scenarios: shallow entry and thin atmosphere, and thick atmosphere and steep entry.

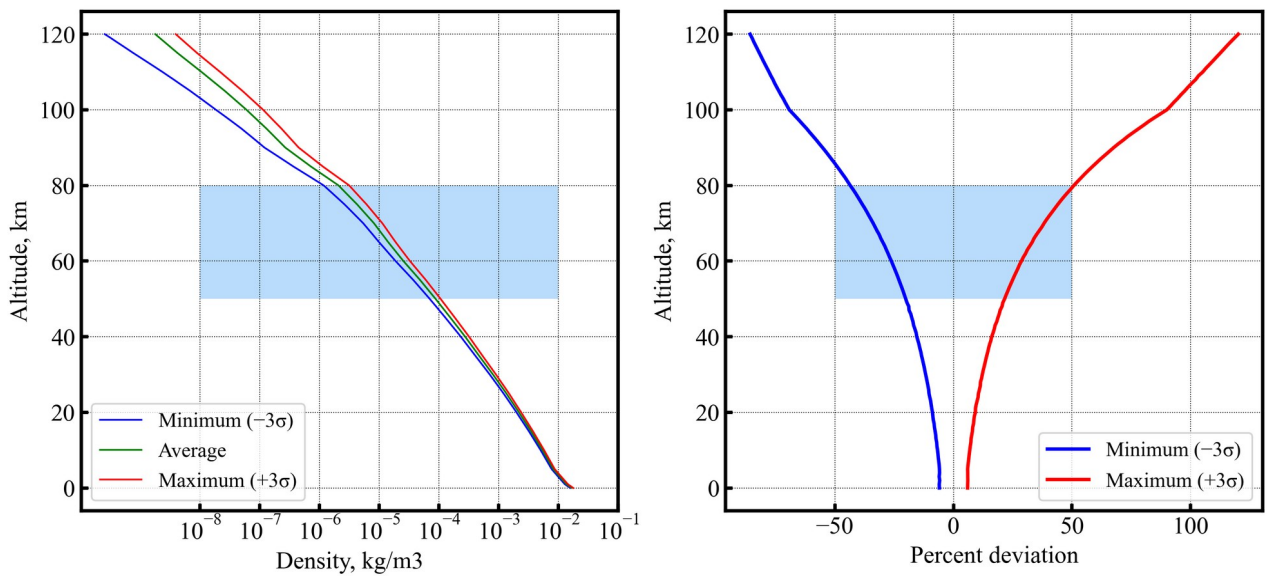

Figure 5. Density profiles from Mars-GRAM (left) and percent deviation from nominal (right).

# VI. TITAN

Saturn's largest moon Titan is the only moon in our Solar System with an atmosphere and makes it unique in several ways such as the only other place with surface liquids [27]. Titan's low gravity and greatly extended thick atmosphere make it the ideal destination for aerocapture, providing the largest corridor width of any destination [28]. Its small size makes it particularly difficult to insert orbiters using conventional propulsion [29, 30]. However, aerocapture is promising alternative for future missions. Aerocapture at has applications for a future Titan orbiter, following the Dragonfly mission, to perform global mapping of the Titan surface and its lakes and seas [31, 32]. Largely due to in-situ date from the Huygens lander, Titan's density profile is well constrained. Figure 6 (left) shows the minimum, average, and maximum (3σ) density profiles from Titan-GRAM. The shaded altitude band of 300–450 km is where most of the deceleration occurs for aerocapture at Titan. Figure 6 (right) shows the percentage deviation (3σ) from the average as a function of altitude. The uncertainty in the density profile increases with altitude, reaches a maximum of about 40% near 100 km above the surface and then decreases. It is not clear this is an artifact of the assumptions used in the model, or indeed a real effect. In the altitude range of 300–450 km km relevant to aerocapture, the expected 3σ density variation is approximately ±30% which is comparable to that at Venus. It is also worth mentioning that though Venus and Titan atmosphere are quite different in terms of their temperature (737K vs 94K) and chemistry ($CO_2$ vs $N_2$), they share several physical similarities, such as both being relatively thick, super-rotating atmospheres with the planetary body rotating slowly and significant greenhouse warming in the lower troposphere.

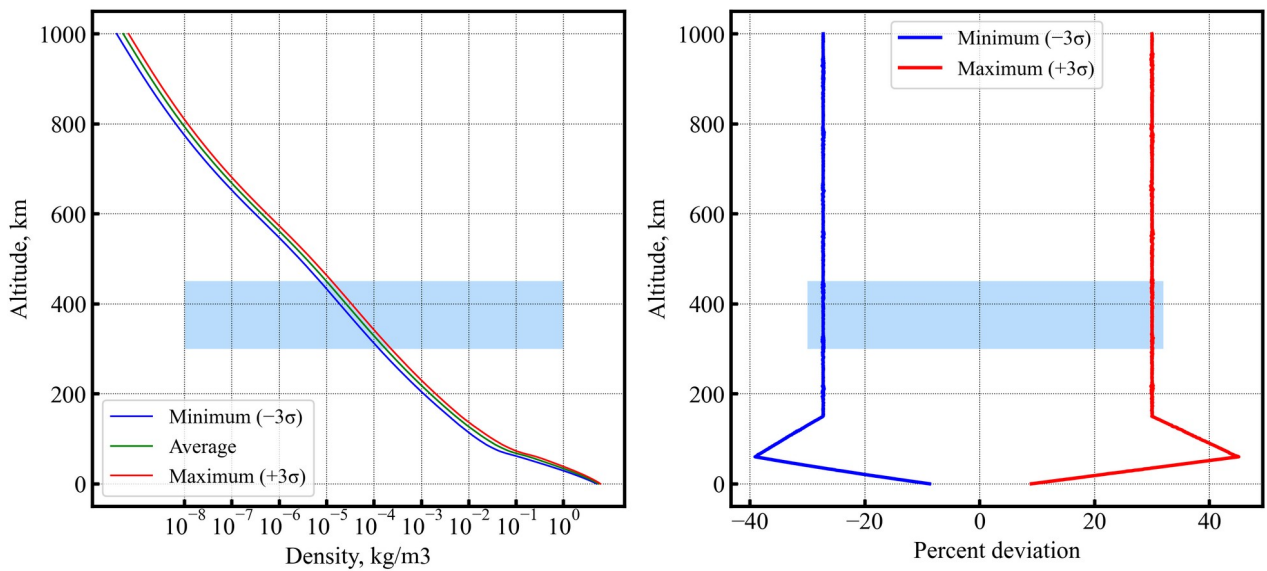

Figure 6. Density profiles from Titan-GRAM (left) and percent deviation from nominal (right).

## VII. URANUS AND NEPTUNE

At the far reaches of the outer Solar System, the ice giants Uranus and Neptune are the last class of planets yet to be explored using orbiter spacecraft. Their enormous heliocentric distance presents significant mission design challenges. The 2023-2032 Planetary Science Decadal Survey has identified a Uranus Orbiter and Probe (UOP) as the top priority for a Flagship mission in the next decade. Even though Uranus and Neptune are both equally compelling scientifically [33], Uranus is less demanding from a mission design perspective with propulsive insertion. Aerocapture was not considered during the Uranus mission studies [34, 35], but aerocapture has been shown to be strongly enhancing to enabling technology for ice giant missions [36]. With aerocapture, both Uranus and Neptune would be equally accessible. Recent studies have shown that aerocapture enables significantly shorter flight times to Uranus than possible with propulsive insertion [37, 38], especially with new high energy launch vehicles [39]. Similar results have been been obtained for Neptune where the performance benefits of aerocapture are even greater [40, 41]. However, the lack of any in-situ measurements presents a challenge as the model predicted atmospheric uncertainties cannot be verified. Figure 7 (left) shows the minimum, average, and maximum (3σ) density profiles. The shaded altitude band of 200–400 km above the 1-bar pressure level is where most of the deceleration occurs at Uranus [42]. Figure 6 (right) shows the percentage deviation (3σ) from the average as a function of altitude. In the altitude range of 200–400 km km relevant to aerocapture, the expected 3σ density variation is approximately ±30%. This must be taken as a "optimistic" estimate until in-situ data becomes available and the actual uncertainty may be much higher.

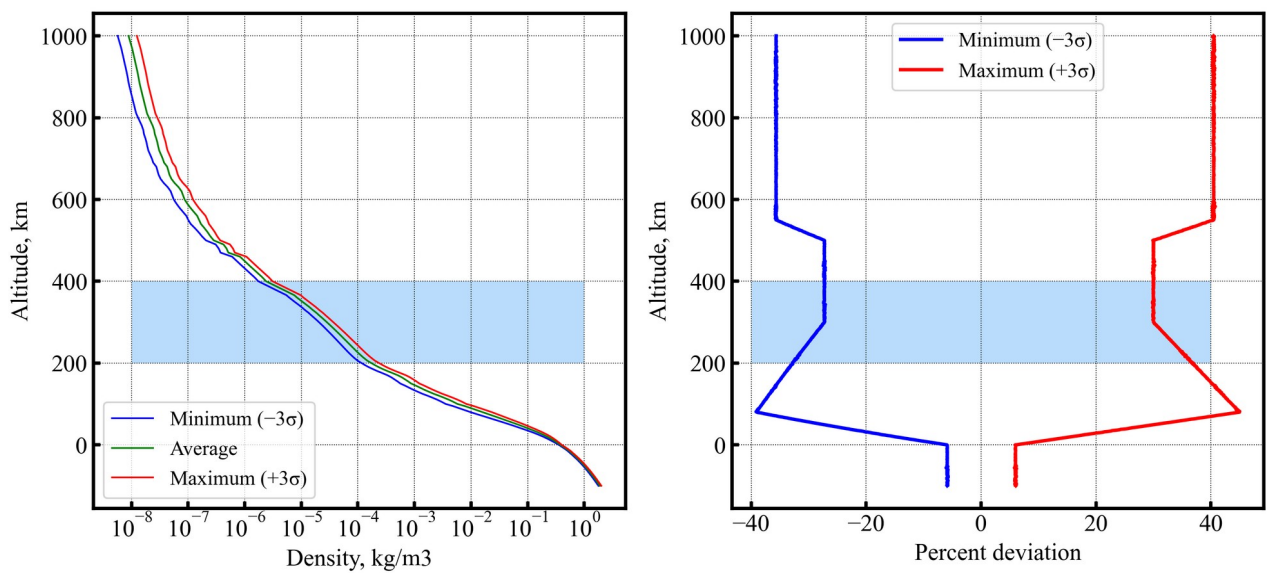

Figure 7. Density profiles from Uranus-GRAM (left) and percent deviation from nominal (right).

Figure 8 (left) shows results from a Neptune-GRAM run. Figure 8 (right) shows the percentage deviation ($3\sigma$) from the average, which is similar to Uranus likely indicating the same uncertainty model is used for both planets. Neptune-GRAM provides a legacy FMINMAX parameter which provides global minimum (-1) and maximum (+1) bounding density profiles which are shown in Figure 9. Considering the full range of FMINMAX, a more "conservative" density variation in the 200–400 km altitude range is (0.1x, 3x) where x indicates a multiplication factor. Until in-situ data from a probe becomes available, the more conservative estimate is recommended.

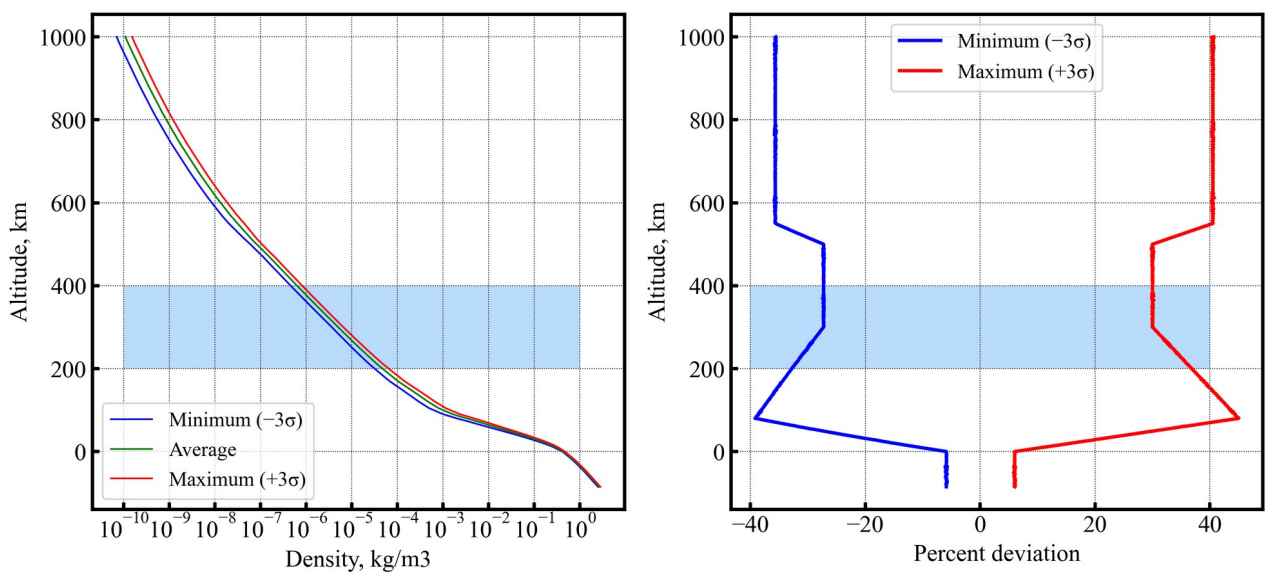

Figure 8. Density profiles from Neptune-GRAM (left) and percent deviation from nominal (right).

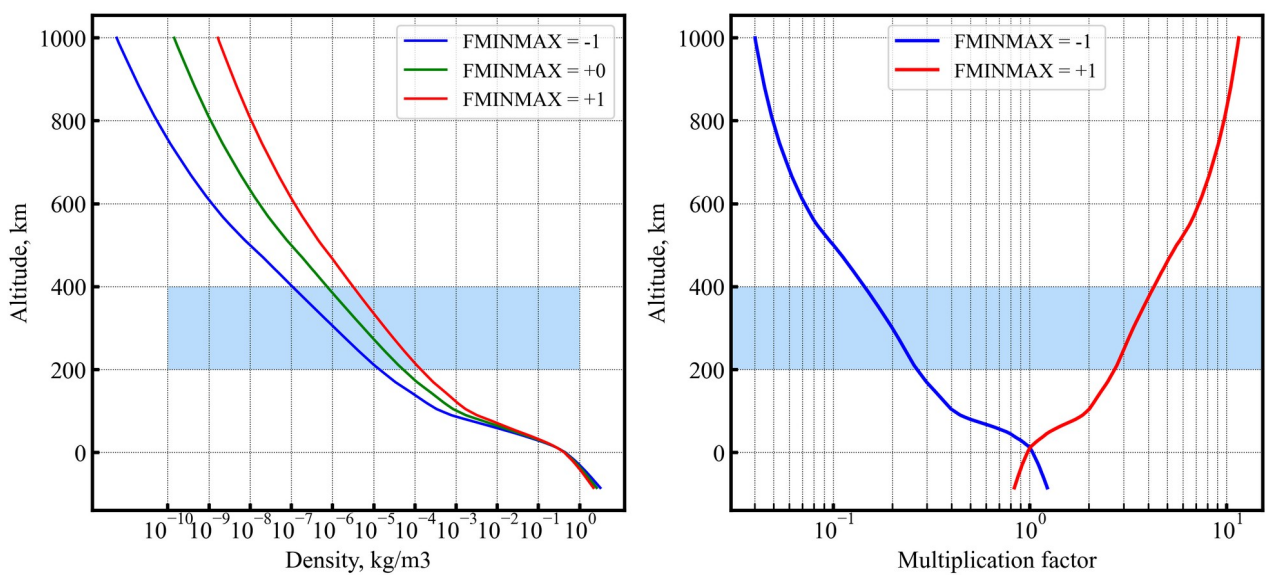

Figure 9. FMINMAX [-1, 0, +1] from Neptune-GRAM (left) and percent deviation (right).

## VIII. DESIGN FOR ATMOSPHERIC UNCERTAINTIES

The aerocapture mission design must account for the expected atmospheric uncertainties to assure the guidance scheme can successfully steer the vehicle to the desired exit state [43]. An important part of the mission design is the selection of the target entry flight path angle (EFPA) for the aerocapture missions. Proper selection of the EFPA can be used to ensure the on-board guidance is given the best possible initial state for targeting the desired exit state. Figure 10 shows an example of the target EFPA selection which accounts for the atmospheric uncertainties for a Uranus mission. The blue, green, and red boxes indicate the aerocapture corridor for minimum, average, and maximum expected density profiles. Note that the corridor changes depending on the atmospheric profile. As the atmosphere becomes thicker, the aerocapture corridor becomes more shallower. For example, if the atmosphere encountered is blue (minimum), then any EFPA shallower than the top boundary (overshoot limit) may result in the vehicle not getting captured. Any EFPA steeper than the bottom boundary (undershoot limit) may result in undershoot of the target orbit or failure to exit altogether. Ideally, the target EFPA is chosen such that two limiting cases are both within the capability of the guidance: a shallow entry and thin atmosphere, and a steep entry and thick atmosphere can both be handled with some additional margin. Graphically, this can be seen by selecting a target EFPA such that both the steep (-3σ) and shallow (+3σ) EFPA values fall within all three boxes. In the case of Uranus example in Figure 10, the corridor even with a L/D=0.24 aeroshell and high entry speed (30 km/s) is only about 1 deg. wide, leaving little margin on either side of the min-density-overshoot or the max-density undershoot. If the corridor width is not high enough to cover the full extent of atmospheric uncertainties, then it is recommended to bias the target EFPA towards the steep side to reduce the high risk of overshoot, at the expense of incurring a small risk of undershoot. In addition, on-board density estimation during the descending leg of aerocapture, and using it during the apoapsis prediction phase of the guidance is critical to ensure mission success in atmospheres with very large uncertainties [44, 45].

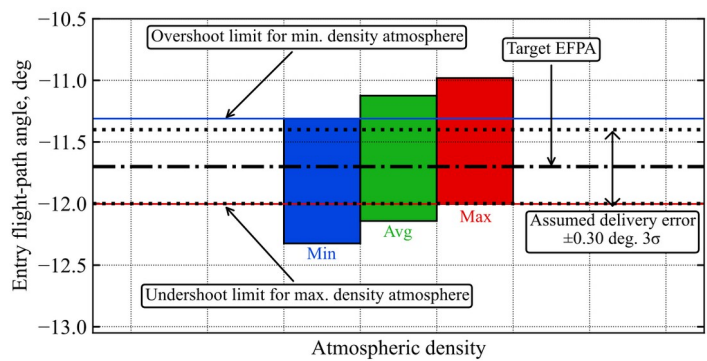

Figure 10. Graphical method for selection of target EFPA incorporating atmospheric uncertainties.

## IX. CONCLUSIONS

The paper used the GRAM suite to perform a comparative study of the atmospheric uncertainties and provided design rules for aerocapture missions. The atmospheres of Venus, Mars, and Titan are well-characterized for engineering purposes, due to the availability of in-situ data. At the altitude ranges relevant for aerocapture, the 3σ density variation is approximately ±30%, ±50%, ±30% for Venus, Mars, and Titan respectively. With no in-situ data, the atmospheres of Uranus and Neptune are not as well characterized. For both Uranus and Neptune, the GRAM suite provides a 3σ density variation of approximately ±30% which must be considered an "optimistic" estimate. Considering the full range of FMINMAX, a more "conservative" density variation in the 200–400 km altitude range is (0.1x, 3x) where x indicates a multiplication factor. Until in-situ data from an atmospheric probe becomes available, the more conservative estimate is recommended as a worst-case scenario for mission concept studies. The study presented a graphical method for selection of the optimal EFPA when considering the atmospheric uncertainties to ensure the on-board guidance is given the best possible initial state for targeting the desired exit state.

## DATA AVAILABILITY

The atmospheric dataset used in the study was created using NASA Global Reference Atmospheric Model (GRAM) Suite v1.5, and is available to download as a zip file at https://doi.org/10.13140/RG.2.2.28166.34880. The data and the code used to make the study results will be made available by the author upon reasonable request.